\documentclass[twocolumn,letter]{jpsj3} 
\usepackage{graphicx}
\usepackage{dcolumn}
\usepackage{bm}
\usepackage{color}

\title{Superconductivity of 2.2 K under Pressure in Helimagnet CrAs}

\author{Hisashi \textsc{Kotegawa}$^{1}$\thanks{E-mail address: kotegawa@crystal.kobe-u.ac.jp}, Shingo \textsc{Nakahara}$^{1}$, Hideki \textsc{Tou}$^{1}$, and Hitoshi \textsc{Sugawara}$^{1}$}

\inst{$^{1}$Department of Physics, Kobe University, Kobe 658-8530, Japan

}

\abst{
We report resistivity measurements of the helimagnet CrAs under pressures. The helimagnetic transition with $T_N \sim 265$ K at ambient pressure is completely suppressed above a critical pressure of $P_c \sim 0.7$ GPa, and superconductivity is observed at $\sim2.2$ K for zero resistance, which exists in a wide pressure range extending beyond 3 GPa. Both the upper critical field $H_{c2}$ and the coefficient $A$ in the resistivity increase toward $P_c$, suggesting that the superconductivity of CrAs is mediated by electronic correlations enhanced in the vicinity of the helimagnetic phase. 
}

\kword{CrAs, superconductivity, helimagnet, pressure}

\begin{document}
\maketitle

$3d$ electron systems can offer stages that induce intriguing superconductivity such as that realized in cuprates, Fe pnictides, and cobalt oxyhydrate.
Superconducting (SC) mechanisms and symmetries are diverse depending on the material, and thus the discovery of superconductivity in a new system is crucial for the development of the field of research on superconductivity.
In this paper, we report the discovery of a pressure-induced superconductivity of $\sim2.2$ K in the helimagnet CrAs through resistivity measurements.
This is the first example of superconductivity in Cr-based magnetic systems.
Very recently, W. Wu and coworkers have independently obtained similar results on the occurrence of superconductivity under high pressure in CrAs through resistivity and susceptibility measurements.\cite{Wu2}

CrAs has an orthorhombic MnP-type crystal structure with the space group of $Pnma$ and shows a first-order magnetic transition at $T_N \sim 265$ K.
Most research studies of the magnetic property of CrAs have been performed in the 1970s.
The magnetic structure of CrAs is a double-helical one represented by a propagation vector of $0.354 \cdot 2\pi c^*$ and a magnetic moment of $\sim1.7\mu_B$/Cr that lies in the $ab$ plane.\cite{Watanabe,Selte}
The magnetic transition is accompanied by a large magnetostriction of $\Delta b/b = +5.5$\%, $\Delta a/a = -0.3$\%, and $\Delta c/c = -0.9 $\% below $T_N$.\cite{Boller,Suzuki}
The crystal structure in the helimagnetic phase has been reported to belong to the same space group as the paramagnetic (PM) phase,\cite{Boller,Suzuki} although the magnetic structure is incommensurate.

Single crystals of CrAs were prepared by the Sn-flux method similar to that described in Ref.~6.
The resistivity measurements under pressure were performed using samples \#1 and \#2, which are from different batches.
Sample \#1 was obtained accidentally from a mixture of Sr:Cr:As:Sn=1:2:2:20, which was prepared with the aim of producing SrCr$_2$As$_2$; however, CrAs was obtained as the main product.
For sample \#2, a mixture of Cr:As:Sn=1:1:10 was prepared.
For both samples, the mixture was placed in an alumina crucible and sealed in an evacuated quartz ampoule.
The ampoule was heated slowly up to 1050 $^\circ$C, and held there for 2 h, and then cooled to 600 $^\circ$C at a rate of -5 $^\circ$C/h.
After centrifuging the flux, the ampoule was cooled to room temperature.
The crystals obtained grew along the $a$-axis.\cite{Laue} 
Electrical resistivity ($\rho$) measurement at high pressures of up to $\sim3$ GPa was carried out using an indenter cell.\cite{indenter}
The current was made to flow along the $a$-axis.
We used Daphne 7474 as the pressure-transmitting medium.\cite{Murata}
The applied pressure was estimated from the $T_{c}$ of the lead manometer; however, it has been reported for Daphne 7474 that the pressure release from room temperature to low temperatures is $\sim0.3$ GPa.\cite{Murata}
To estimate the pressure at high $T_N$, we assumed that pressure is unchanged below 100 K and depends on temperature linearly from 100 K to room temperature, increasing at a rate of $1.5 \times 10^{-3}$ GPa/K.

\begin{figure}[htb]
\centering
\includegraphics[width=0.95\linewidth]{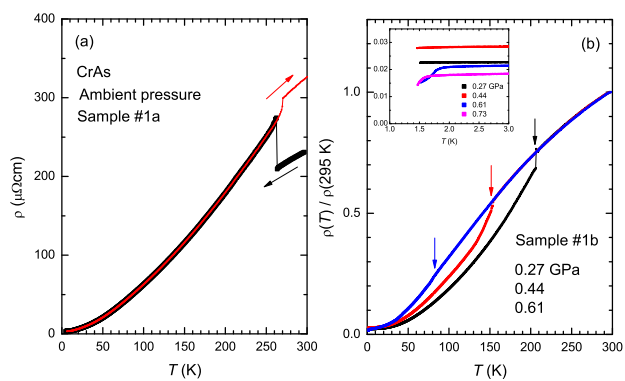}
\caption[]{(color online) (a) Temperature dependence of resistivity of CrAs at ambient pressure. A clear anomaly was observed at $T_N$ accompanied by hysteresis, but the resistivity does not return to its original value, indicating that cracks are induced in the crystal owing to the large magnetostriction. This behavior was observed regardless of the difference between the samples \#1 and \#2. (b) Temperature dependence of resistivity of sample \#1 at several pressures. $T_N$ is strongly suppressed under pressure. The inset shows the resistivity at low temperatures.
}
\end{figure}

Figure 1(a) shows the temperature dependence of resistivity for CrAs measured at ambient pressure.
A discrete jump is observed below 263 K on cooling; however, the drop in $\rho$, which shows an opposite behavior to that shown in Fig.~1(a), is expected to be intrinsic.\cite{Wu}
In this setting, the resistivity does not return to its original value at room temperature, as shown by the red curve.
This indicates that cracks are induced in the crystal owing to the large magnetostriction.
We measured different crystals several times for both samples \#1 and \#2, but a similar behavior was again observed.
In fact, crystals often collapse, once they enter into the ordering state.
The precise estimation of the residual resistivity ratio (RRR) $=\rho({\rm 300 \ K})/\rho_0$ at ambient pressure is difficult, because the resistivity does not return to its original value at room temperature.
If we define RRR$_{T_N}=\rho(T_N^-)/\rho_0$, where $\rho(T_N^-)$ is the resistivity just below $T_N$, RRR$_{T_N}$ is not affected significantly by the repetition of magnetic transition. 
However, if we evaluate the residual resistivity $\rho_0$ at ambient pressure, the crystal experiences a magnetic transition at least once.
This means that the $\rho_0$ at ambient pressure is inevitably evaluated for crystals with cracks.
It is unclear whether damage to crystals extends up to the microscopic level.

Figure 1(b) shows the temperature dependence of resistivity of sample \#1, measured at several pressures, where the resistivity is normalized by the value at 295 K.
The resistivity estimated actually increases by 1.8 times from ambient pressure to 0.73 GPa owing to the occurrence of cracks.
$T_N$ apparently decreases with increasing pressure, and the anomaly at $T_N$ is not observed above the critical pressure $P_c\sim0.7$ GPa, which is consistent with the earlier result obtained using a polycrystalline sample\cite{Zavadskil} as well as with a recent report.\cite{Wu2}
The hysteresis at $T_N$ was observed below $\sim0.7$ GPa; thus, it is conjectured that the magnetic transition is of the first order even near $P_c$.
The inset shows the resistivity at low temperatures, where a signature of superconductivity is observed; however, zero resistance was not achieved in the measured temperature range.
The RRR for the present sample (\#1) was estimated to be $\sim60$ at 0.73 GPa just above $P_c$, where the resistivity returns to its original value at room temperature after the measurement at low temperatures.

\begin{figure}[htb]
\centering
\includegraphics[width=0.95\linewidth]{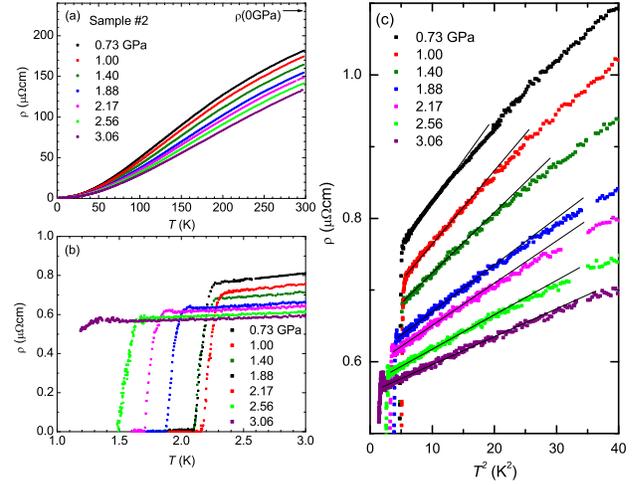}
\caption[]{(color online) (a) Temperature dependence of resistivity of CrAs (sample \#2) at several pressures in the PM phase. This crystal does not experience the ordering state. RRR is estimated to be $\sim290$ at 0.73 GPa. The arrow indicates the resistivity at ambient pressure and room temperature. (b) Resistivity at low temperatures. A clear SC transition is observed with a maximum $T_c$ of 2.17 K at 1.00 GPa. (c) $T^2$ vs $\rho$ plot. The slope decreases gradually with increasing pressure. The data obeys the FL form only in the narrow temperature range.
}
\end{figure}

Figure 2(a) shows the temperature dependence of the resistivity of CrAs (sample \#2) under a different setting.
Here, we applied pressure beyond $P_c$ at room temperature as a first step to avoid damage to the sample, so that this crystal does not experience the helimagnetic transition.
The RRR at 0.73 GPa was estimated to be $\sim290$.
It is unclear whether the original sample quality or the experience of the magnetic transition induces the difference in RRR between samples \#1 and \#2.
Figure 2(b) shows the resistivity at low temperatures, measured after performing this procedure.
A clear SC transition can be observed, and the maximum $T_c$ of $2.17$ K for zero resistance is attained at 1.00 GPa.
The $T_c$ slightly increases from 0.73 GPa close to $P_c$, reaches its maximum at approximately 1.00 GPa, and decreases with increasing pressure.
At 3.06 GPa, the onset of superconductivity can be confirmed in the measured temperature range.
The $T_c$ for zero resistance is significantly higher than that for sample \#1, where the onset of superconductivity is observed below $\sim1.6$ K at 0.73 GPa, as shown in the inset of Fig.~1(b).
Figure 2(c) shows the $T^2$ vs $\rho$ plot. 
The resistivity obeys the Fermi liquid (FL) form of $\rho(T)=\rho_0 + AT^2$ in a narrow temperature range, and its temperature range is likely to extend slightly under higher pressures.
The slope corresponds to the coefficient $A$, which represents the inelastic scattering between electrons and is generally proportional to the square of the effective electron mass.
$A$ clearly decreases with increasing pressure.

\begin{figure}[htb]
\centering
\includegraphics[width=0.7\linewidth]{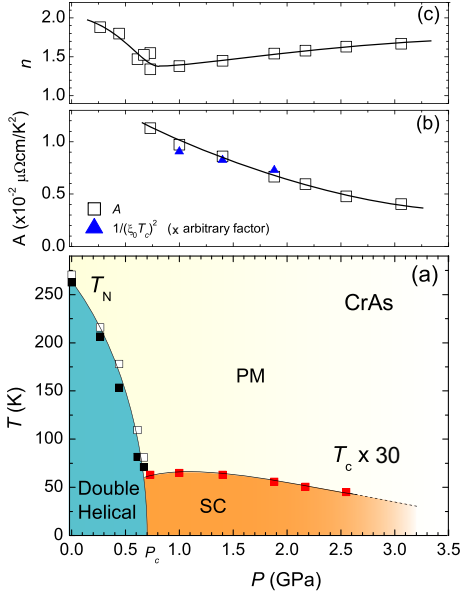}
\caption[]{(color online) (a) Pressure-temperature phase diagram of CrAs. The helimagnetic phase disappears above $P_c\sim0.7$ GPa, and superconductivity appears in the PM phase. The closed (open) squares represent $T_N$ obtained during cooling (warming). The observation of hysteresis up to $P_c$ indicates that helimagnetic-paramagnetic transition is of the first order even under pressure.  (b) The pressure dependence of the coefficient $A$, which is estimated from the resistivity below 4 K. Above $P_c$, $A$ decreases gradually with increasing pressure. The estimated $1/(\xi_0T_c)^2$ is also plotted, showing good scaling with $A$.  (c) Pressure dependence of $n$ estimated from the resistivity between $T_c$ and 10 K. The clear deviation from the FL behavior is confirmed. The highest $T_c$ of 2.17 K is achieved at 1.00 GPa, accompanied by the increase in $A$ and the deviation from the FL behavior.
}
\end{figure}

The pressure-temperature phase diagram of CrAs up to $\sim3$ GPa is shown in Fig.~3(a).
The helimagnetic phase disappears above $P_c\sim0.7$ GPa.
The SC phase appears in the PM state, extending to relatively high pressures.
Figures 3(b) and 3(c) show the pressure dependence of $A$, which is estimated using $\rho(T)=\rho_0 + AT^2$ for the data between $T_c$ and 4 K, and the pressure dependence of $n$, which is evaluated using $\rho(T)=\rho_0 + A'T^n$ between $T_c$ and 10 K.
$A$ decreases gradually with increasing pressure, indicating that the effective electron mass is enhanced by magnetic correlations near $P_c$.
$n$ is close to 2 in the low-pressure region and has its minimum just above $P_c$, followed by a gradual increase under higher pressures.
We obtained $n=1.4\pm0.1$ near $P_c$, which is close to $n=1.5\pm0.1$ shown by Wu {\it et al}.\cite{Wu2}
The highest $T_c$ of 2.17 K is achieved at 1.00 GPa, accompanied by the increase in $A$ and the deviation from the FL behavior.
This behavior is reminiscent of a typical case of heavy-fermion superconductors,\cite{Demuer} which have a quantum critical point of magnetic origin.
It is conjectured that magnetic fluctuations play a vital role in inducing superconductivity in CrAs.
However, the difference of CrAs from heavy-fermion systems is the first-order separation of the magnetic and PM phases in CrAs.
This situation is reminiscent of the Fe-based superconductor SrFe$_2$As$_2$, where a $T_c$ of 34 K is achieved in the vicinity of the magnetic phase, but the magnetic and PM phases are separated by a first-order transition.\cite{Kotegawa}
In this case, the coexistence of magnetism and superconductivity is not simple, and the hybrid state of both phases has been reported in SrFe$_2$As$_2$.\cite{Kitagawa}
It is an intriguing issue whether superconductivity can coexist with the helimagnetic state in CrAs.
Another interesting similarity to Fe-based superconductors is that the magnetic transition is accompanied by the structural phase transition.
The contribution of orbital fluctuations to superconductivity might be an interesting issue to be considered in CrAs.

Wu {\it et al.} have reported zero resistance below $\sim1.5$ K,\cite{Wu2} whereas we observed it below $\sim2.2$ K.
The RRR range of their samples is $240-327$, and thus there is no significant difference from that of our sample \#2.
The pressure-transmitting media used are different from each other, but we consider that the properties of the media used, i.e., Daphne 7474 (our study) and glycerol (Wu {\it et al.}'s study) are similar.
A possible reason for this is the experience of the magnetic transition with a large magnetostriction, although we do not know the experimental procedure used by Wu {\it et al.}
A systematic study is required to clarify the cause of the difference in $T_c$.

\begin{figure}[htb]
\centering
\includegraphics[width=0.95\linewidth]{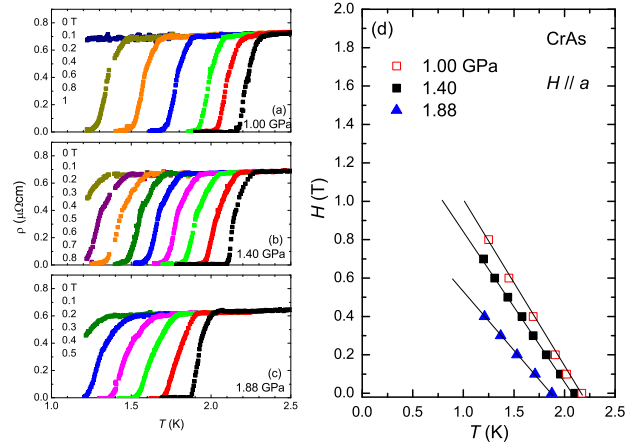}
\caption[]{(color online) (a-c) Temperature dependences of the resistivity under different magnetic fields along the $a$-axis. At 1.88 GPa, superconductivity is suppressed by the small magnetic field. (d) Temperature dependence of $H_{c2}$. The magnitude of the slope decreases with increasing pressure. From the slopes, we obtained the orbital field $H_{c2}^{orb}(0)$ to be 1.37 T (1.00 GPa), 1.16 T (1.40 GPa), and 0.79 T (1.88 GPa).
}
\end{figure}

Figures 4(a)-4(c) show the temperature dependence of the resistivity under different magnetic fields along the $a$-axis.
For 1.00, 1.40, and 1.88 GPa, the field dependences of $T_c$ are obtained and displayed in Fig.~4(d).
The initial slope of $(-dH_{c2}/dT)_{Tc}$ gives the orbital field $H_{c2}^{orb}$ through $H_{c2}^{orb}(0) = 0.727 (-dH_{c2}/dT)_{Tc} T_c$ in the clean limit.\cite{Helfand}
The slopes are estimated to be 0.91 T/K (1.00 GPa), 0.80 T/K (1.40 GPa), and 0.62 T/K (1.88 GPa), giving $H_{c2}^{orb}(0)$ of 1.44 T (1.00 GPa), 1.22 T (1.40 GPa), and 0.85 T (1.88 GPa), respectively.
These values correspond to the Ginzburg-Landau coherence length $\xi_0$, through $H_{c2}^{orb}(0) = \Phi_0/2\pi \xi_0^2$ where $\Phi_0$ is the quantum fluxoid.
$\xi_0$ along the $bc$ plane is estimated to be 151 \AA \ (1.00 GPa), 164 \AA \ (1.40 GPa), and 197 \AA \ (1.88 GPa).

We perform a simple comparison between the pressure dependences of $H_{c2}$ and $A$, which corresponds to the effective electron mass.
$\xi_0$ is given by $\xi_0 \simeq \hbar v_F/\pi \Delta_0 = \hbar^2 k_F / \pi \Delta_0 m^*$, where $v_F$, $k_F$, $\Delta_0$, and $m^*$ are Fermi velocity, Fermi wave number, SC gap size, and effective electron mass, respectively.
If we assume for simplicity that $k_F$ and $\Delta_0/k_BT_c$ are independent of pressure, we obtain the relation $A \propto 1/(\xi_0T_c)^2$, which is plotted in Fig.~3(b).
The good scaling between $A$ and $1/(\xi_0T_c)^2$ strongly suggests that the superconductivity in CrAs is mediated by electronic correlations enhanced near the helimagnetic phase.
Thus, the elucidations of the underlying correlations in the PM phase and the pressure evolution of the magnetic structure are important for understanding superconductivity in CrAs.

In summary, we have completed the pressure-temperature phase diagram of CrAs up to $\sim3$ GPa.
The helimagnetic phase is suppressed at $\sim0.7$ GPa, and superconductivity is realized in a wide pressure range in the PM state with a maximum $T_c\sim2.2$ K at 1.0 GPa.
The experimental results indicate that the superconductivity of CrAs is likely to be sensitive to the sample quality and experimental conditions.
The pressure dependences of $A$ and $H_{c2}$ suggest that superconductivity is mediated by electronic correlations enhanced in the vicinity of the helimagnetic phase.
The phase diagram of and the behavior observed in CrAs have similarities to those of heavy-fermion superconductors and some Fe-based superconductors.
CrAs is an interesting material for investigating how SC symmetry is realized in the vicinity of the helimagnetic phase with a large magnetostriction.

\section*{Acknowledgements}

We thank Hisatomo Harima for helpful discussions.
This work has been partially supported by Grants-in-Aid for Scientific Research (Nos. 22340102, 20102005, and 24340085) from the Ministry of Education, Culture, Sports, Science and Technology (MEXT) of Japan.


\begin{thebibliography}{99}
\bibitem{Laue}
In Ref.~6, it has been described that the crystal is grown along the $b$-axis. In our Laue measurement, the crystal was grown along the $a$-axis.
\end{thebibliography}
\end{document}